\documentclass[conference]{IEEEtran}
\IEEEoverridecommandlockouts

\usepackage{amsmath,amssymb,amsfonts}
\usepackage{graphicx}
\usepackage{textcomp}
\usepackage{xcolor}
\usepackage{booktabs}
\usepackage{color}
\usepackage{ mathrsfs }
\usepackage{float}
\usepackage{multirow}
\usepackage{transparent}
\usepackage{subcaption}
\usepackage{siunitx}
\usepackage{comment}
\usepackage{caption}
\usepackage[bookmarks=false]{hyperref}
\usepackage{lipsum} 
\usepackage{fancyhdr}

\def\BibTeX{{\rm B\kern-.05em{\sc i\kern-.025em b}\kern-.08em
    T\kern-.1667em\lower.7ex\hbox{E}\kern-.125emX}}

\begin{document}

\title{AdaSense: Adaptive Low-Power Sensing and Activity Recognition for Wearable Devices}

\author{\IEEEauthorblockN{Marina Neseem}
\IEEEauthorblockA{\textit{School of Engineering} \\
\textit{Brown University}\\
Providence, RI \\
marina\_neseem@brown.edu}
\and
\IEEEauthorblockN{Jon Nelson}
\IEEEauthorblockA{\textit{School of Engineering} \\
\textit{Brown University}\\
Providence, RI \\
jon\_nelson1@brown.edu}
\and
\IEEEauthorblockN{Sherief Reda}
\IEEEauthorblockA{\textit{School of Engineering} \\
\textit{Brown University}\\
Providence, RI \\
sherief\_reda@brown.edu}
}

\IEEEaftertitletext{\vspace{-2\baselineskip}}

\maketitle

\thispagestyle{fancy}
\cfoot{\thepage}

\begin{abstract}
Wearable devices have strict power and memory limitations. As a result, there is a need to optimize the power consumption on those devices without sacrificing the accuracy. This paper presents AdaSense: a sensing, feature extraction and classification co-optimized framework for Human Activity Recognition. The proposed techniques reduce the power consumption by dynamically switching among different sensor configurations as a function of the user activity. The framework selects configurations that represent the pareto-frontier of the accuracy and energy trade-off. AdaSense also uses low-overhead processing and classification methodologies. The introduced approach achieves 69\% reduction in the power consumption of the sensor with less than 1.5\% decrease in the activity recognition accuracy.
\end{abstract}


\section{\textbf{Introduction}}
\label{sec:intro}

Recent advances in wearable devices show a strong potential to revolutionize applications in both health monitoring and patient diagnosis. 
For example, wearable devices have a big influence in post-op rehabilitation. A successful surgery depends on monitoring the patient’s condition after the surgery, which is carried out through clinical visits. However, these visits are often insufficient and can be greatly enhanced by continuous monitoring using wearable devices. For example, it was recently shown that a wearable stretch sensor can be used to provide feedback for physical therapy or rehabilitation exercises \cite{skin_stretch_sensor,rehabilitation}. Also, special robotic devices can be used for post-stroke shoulder rehabilitation to identify misalignment \cite{ortho}. Furthermore, a recent work uses wearable devices to detect hidden anxiety and depression in young children \cite{mcginnis2019rapid}.
Activity recognition models have also been used to enhance health monitoring for the elderly. A case study uses wearable sensors and other environmentally placed sensors to predict health decline and critical health situations \cite{elder_1} and another one uses inertial wearable sensors and gait detection to provide useful digital bio-markers in dementia \cite{dimentia}.

Wearable devices have very limited power and memory constraints. For example, Human Activity Recognition (HAR) platforms need to reach a satisfying activity recognition accuracy while consuming low power, resulting in a power-accuracy trade-off, which creates the need to co-optimize all the power consuming components of the device. 

Usually wearable devices have two main energy consuming components: the sensors and the processing unit. In HAR platforms \cite{bhat2018online} \cite{8782506}, there are two main stages: first, collecting real time data from an accelerometer and/or a gyroscope, and second, pre-processing the data to extract some features before forwarding them to a classifier to analyze the user activity, such as walking, sitting, running, etc. Many recent works have presented optimized classification algorithms that are low in power and memory \cite{kumar2017resource,gupta2017protonn,kusupati2018fastgrnn}. A few recent papers addressed methods to reduce the power consumption of the sensors by shutting down some of them \cite{bhat2019reap} or by switching the sensors to a lower sampling frequencies with less intense user activities \cite{8782506}. However, we observe that the {\it averaging window} is also a key factor affecting the sensor's energy consumption. Moreover, switching the sensors to different sampling rates introduces a classification challenge as each sampling rate provides a feature set of different size; this challenge was overlooked in previous works by retraining different classifiers for each used sampling rate creating a memory overhead. To tackle this challenge, we are the first to consider manipulating the feature extraction step to output the same feature size for data from different sensor configurations.

In AdaSense, we propose a novel low-power sensing technique that optimizes the power consumption of wearables while maintaining high activity recognition accuracy. We co-optimize the sensor, the feature extraction and the classifier to enhance the energy consumption of wearable devices. Our main contributions can be summarized as follows:
\begin{itemize}
    \item \textbf{Sensor Configurations Design Space Exploration:} We provide a complete evaluation for the trade-off between activity recognition accuracy and power consumption under 16 different accelerometer sampling frequency and averaging window combinations to pick the optimal sensor configurations that trade off the accuracy with the power consumption in an optimal way.
    
    \item \textbf{Adaptive Low-Power Sensing technique:} Using the outcomes of the sensor configurations design space exploration, we introduce a novel technique that dynamically switches the sensor operation among these different configurations as a function of the user activity. We also provide detailed results for the recognition accuracy, memory and power consumption in comparison to the previously used techniques.
    
    \item \textbf{Features Extraction and Classification:} We propose a new feature extraction methodology which unifies the features set for heterogeneous data coming from various accelerometer configurations (i.e. sampling frequency and averaging window). This enables the usage of a single classifier that is capable of recognizing the human activity regardless the chosen configuration.
\end{itemize}

The rest of the paper is organized as follows. We review the related work in Section~\ref{sec:previous_work}. Then, we present our activity recognition framework in Section~\ref{sec:recognition_techniques}. In Section~\ref{sec:sensing_techniques}, we analyze the design space of the different sensor configurations, and we propose our adaptive low-power sensing technique. Next, we show the experimental setup and results in Section~\ref{sec:experiments}. Finally, we conclude in Section~\ref{sec:conclusion}.

\section{\textbf{Previous Work}}
\label{sec:previous_work}

Due to the tight energy constraints of wearable devices, recent work have focused on approaches to minimize power consumption while maintaining high performance (recognition accuracy) \cite{8782506}. Researchers have determined that the sensor is responsible for a large fraction of the device's power consumption, reaching about 47\% \cite{bhat2019reap}. Therefore, reducing the sensor's power consumption significantly decreases the total energy requirement for the device. This can be done by switching the sensor to low-power mode, reducing the sampling frequency \cite{8782506}, turning off the sensor intermittently, shutting down one of the sensor's axes \cite{bhat2019reap} or using compressed sensing techniques for selective sampling \cite{gazelle}. However, to the best of our knowledge there has not been any approach that considers the averaging window of the sensor to maximize power savings. 

There are some approaches that consider using approximate circuits to achieve significant power savings without sacrificing much accuracy \cite{approx_divider,2840878_drum}, while others focus on minimizing the power consumption of the used machine learning algorithms \cite{kumar2017resource,gupta2017protonn}. Feature extraction and data processing also require significant energy, so researchers have developed techniques to reduce the complexity of the extracted features in order to save power \cite{feat_ext}. For example, statistical features are relatively simple to calculate, while the Fourier Transform and Discrete Wavelet Transform are more computationally complex, so we can dynamically choose which features to calculate based on the required power budget \cite{bhat2019reap}.

Once the relevant design points have been identified, various approaches attempt to determine the optimal strategy for switching between these design points at runtime. NK {\it et al}. \cite{8782506} propose to switch to power saving design points when the user is doing low-intensity activities because these do not require as many data points to classify. Liu {\it et al}. \cite{7058994} instead propose to use compressed sensing techniques to determine how many samples are needed for reconstruction and consequently sample as needed.  

Finally, the machine learning classifier must be compatible with the dynamic sensor data. If the sensor data is acquired under settings that are different from the training data, then the accuracy can be significantly degraded. NK {\it et al}. \cite{8782506} address this problem by retraining a separate neural network for each design point, while Liu {\it et al}. \cite{7058994} use linear interpolation to normalize for variable sampling rate.

\section{\textbf{Proposed Human Activity Recognition Framework}}
\label{sec:recognition_techniques}

The ultimate goal of our HAR framework is to read an accelerometer's 3-axis data, and analyze them into one of six daily activities: sit, stand, walk, go upstairs, go downstairs and lie down. Two main components are needed to do that task: the feature extraction and the classification. Both tasks should be done on the wearable device, so the needed processing power and storage memory should be meticulously considered. Moreover as mentioned in Section~\ref{sec:intro}, the HAR classifier should handle heterogeneous sensor data of different nature, e.g., different sampling frequencies and averaging windows, without creating a processing or a memory overhead. AdaSense tackles this heterogeneous data problem by using a new feature extraction technique that unifies the size of the features vector regardless the accelerometer's configurations. In this section, we first explain the main components of the HAR framework, then we present our new methodology for feature extraction as well as the used activity classification model.

\begin{figure}[t]
\setlength{\belowcaptionskip}{-15pt}

\centerline{\includegraphics[clip, trim=10pt 20pt 10pt 20pt, width=220pt,height=220pt,keepaspectratio]{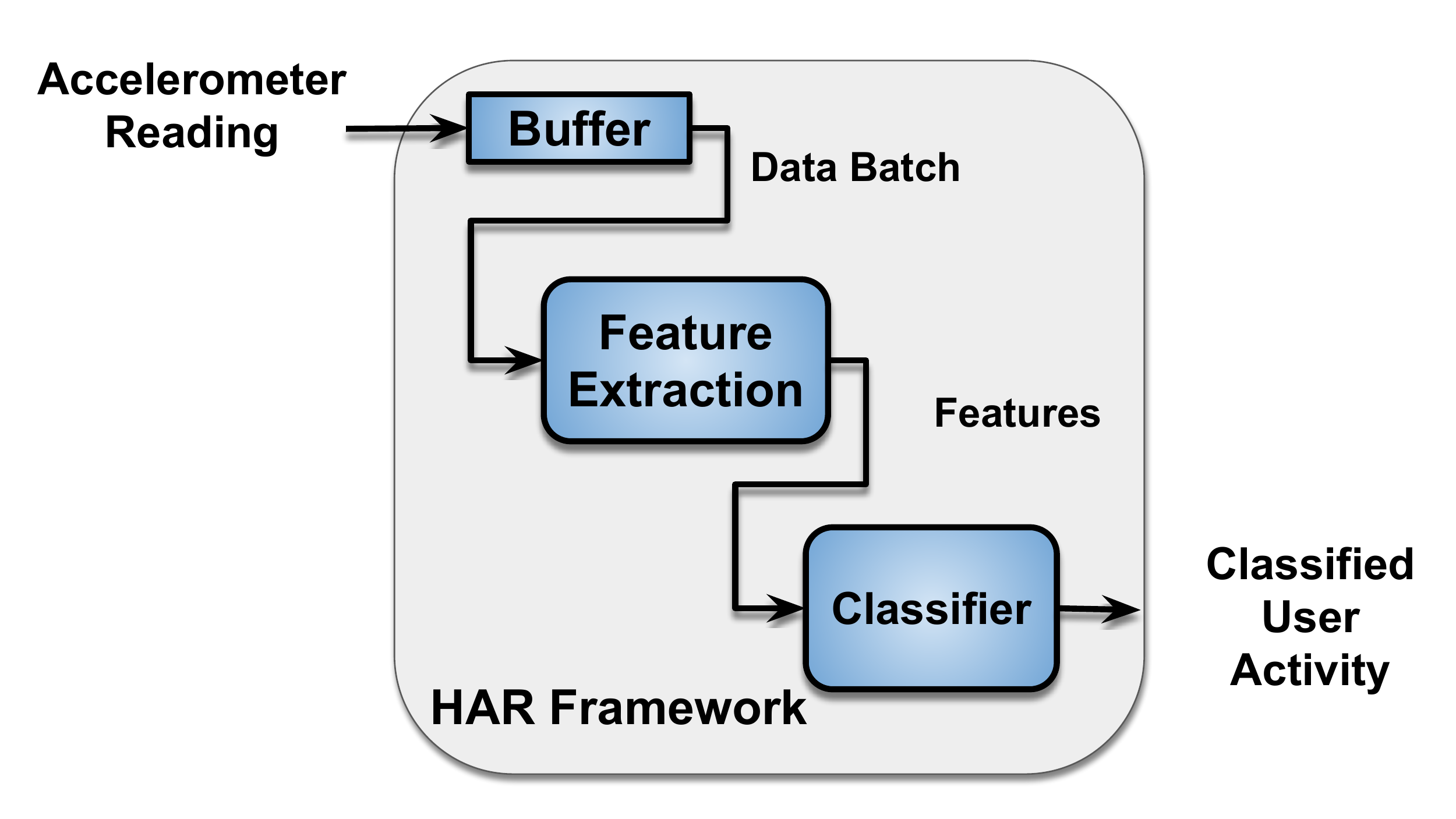}}
\caption{Human Activity Recognition Framework.}
\label{activity_recognition_framework}
\end{figure}

\subsection{\textbf{Activity Recognition Framework Main Components}}
The main components of our HAR framework can be summarized in Fig.~\ref{activity_recognition_framework}. The input is the data collected by the accelerometer, and the output is the user activity class. A batch of sensor data is needed to perform meaningful classification, so a buffer is added to control the classification frequency. The buffer stores the accelerometer data over two consecutive seconds. Then every one second, we push the collected data batch in the buffer through the rest of the pipeline, where we run feature extraction and classification to predict the user activity. We introduce one second overlap between the data batches to give the classifier some insightful information about the previous classification data batch. The challenge is that the data batch size depends on the sampling frequency, i.e., the number of samples stored in the buffer during one second period would be 100 with 100 Hz sampling frequency, and 50 when the sampling frequency is 50 Hz, making the job harder for the classifier. This issue would be handled during the feature extraction by constructing similar size feature set regardless the size of the processed data batch.

\subsection{\textbf{Feature Extraction}}
We can split the used features set into two categories: {\it statistical features} and {\it Fourier transform coefficients}. The statistical features include the mean and the standard deviation of the signal for the x, y and z coordinates. These statistical quantities capture the general structure of the accelerometer data over the selected batch. The Fourier transform coefficients capture frequency information for each activity. However, we noticed that we do not need to use all the Fourier transform components. For the sake of activity recognition, the first three coefficients in each coordinate, representing the frequency components up to 3 Hz, are enough to get around 97\% recognition accuracy. The big advantage of using these features is that the feature vector size would be the same regardless of the size of the processed data batch. This solves part of the classifier's problem; however, the classifier still needs to adapt to the information in the features depending on the used sampling frequency and averaging window.

\subsection{\textbf{Human Activity Classifier}}
AdaSense's classifier must gracefully handle heterogeneous sensor data in order to maintain high accuracy. This task is not trivial because classification accuracy can degrade significantly if the sensor configurations of the test data are different from the configurations of the training data. A commonly used approach is to retrain a different model for each sensor configuration, which is guaranteed to provide high recognition accuracy but adds memory overhead to store multiple classifiers \cite{8782506}. However, since we have unified the size of the feature set for different sampling rates, we can do the classification using one neural network with two layers: one hidden layer with RELU activation function and an output layer with 6 neurons and a softmax. By training this network on data from different sampling frequencies and averaging windows, we expect it to perform well using much less memory depending on the different sensor configurations used, i.e if there are four different sensors configurations, using our method will need one network instead of four different networks using $4\times$ less memory to store the weights.

\section{\textbf{Proposed Low-Power Sensing}}
\label{sec:sensing_techniques}

Wearable devices have two main power consuming components: the sensors and the processing unit. In Section \ref{sec:recognition_techniques}, we described our HAR framework structure and the used methodologies to perform efficient processing and classification. In this section, we describe how AdaSense optimizes the power consumption of the sensor. We first analyze the accuracy and power trade-off for different sensor configurations. Next, we use the outcome of this analysis to design an adaptive controller that changes the sensor configurations according to the user activity.

\subsection{\textbf{Sensor operation Modes}}
Usually sensors has two operation modes: {\it normal mode} and {\it low-power mode}. For the sensor to give less noisy readings, the output is not just the instantaneous reading at the required sampling point. Instead, it is the average of the collected samples over a certain window before the sampling point; the size of that window is called the {\it averaging window}. In normal mode, the sensor stays on all the time, so the averaging window does not affect the power consumption. However, in low-power mode, the sensor switches between normal and suspend modes, so both the sampling frequency and the averaging window determine the time in which the sensor has to be on; hence, significantly affecting the power consumption.

\newcommand{\tabitem}{~~\llap{\textbullet}~~}

\begin{table}
  \centering
  \caption{Acceleromenter Sampling Frequency and Averaging Window Combinations.}
  \begin{tabular}{lll}
    \toprule
    \tabitem  100Hz / 128 (F100\_A128) & \tabitem 50Hz / 128 (F50\_A128) \\
    \tabitem 25Hz / 128 (F25\_A128) & \tabitem 12.5Hz / 128 (F12.5\_A128) \\
    \tabitem 6.25Hz / 128 (F6.25\_A128) & \tabitem 25Hz / 32 (F25\_A32) \\
    \tabitem 12.5Hz / 32 (F12.5\_A32) & \tabitem 6.25Hz / 32  (F6.25\_A32) \\
    \tabitem 50Hz / 16 (F50\_A16) & \tabitem 25Hz / 16 (F25\_A16) \\
    \tabitem 12.5Hz / 16 (F12.5\_A16) & \tabitem 6.25Hz / 16 (F6.25\_A16) \\
    \tabitem 50Hz / 8 (F50\_A8) & \tabitem 25Hz / 8 (F25\_A8) \\
    \tabitem 12.5Hz / 8 (F12.5\_A8) & \tabitem 6.25Hz / 8 (F6.25\_A8) \\
    \bottomrule
  \end{tabular}
  \vspace{-18pt}
\label{Freq_AVG_combinations}
\end{table}

\subsection{\textbf{Sensor configurations Design Space Exploration}}
\begin{figure}[b]
\vspace{-10pt}
\setlength{\belowcaptionskip}{-10pt}
\centerline{\includegraphics[clip, trim=10pt 20pt 10pt 75pt, width=250pt,height=250pt,keepaspectratio]{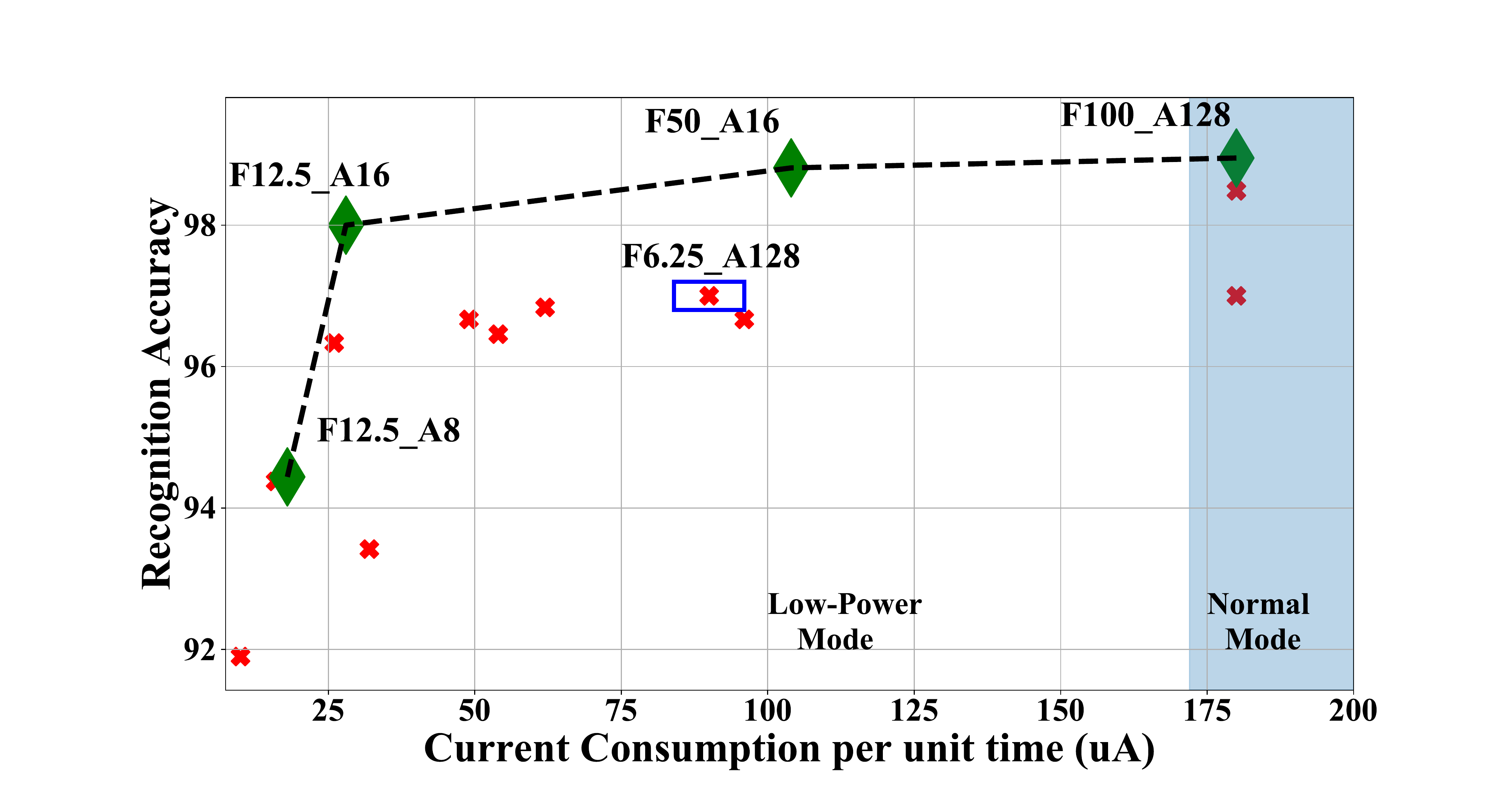}}
\caption{Accelerometer configurations accuracy and power trade-off.}
\label{pareto}
\end{figure}

Using the setup mentioned in Section \ref{sec:recognition_techniques}, we study the power consumption and the recognition accuracy at different sensor configurations. We choose 16 different sampling frequency and averaging window combinations given in Table \ref{Freq_AVG_combinations}, and we analyze the accuracy and power trade-off as illustrated in Fig.~\ref{pareto}. Each point in the graph represents a different sensor configuration, and each configuration has a unique current-accuracy value which is a factor of the sampling rate and the noise due to using lower averaging windows. We observed that the four configurations \emph{\{F100\_A128, F50\_A16, F12.5\_A16, F12.5\_A8\}}, highlighted by the diamond shape, dominate the others, and create a Pareto front that optimally trades off accuracy with power. \emph{F100\_A128} has the highest accuracy and current consumption, whereas \emph{F12.5\_A8} has the lowest accuracy and current consumption. However, the other red points do not offer any benefits in the energy-accuracy trade-off; for example, the point \emph{F6.25\_A128} marked by the blue rectangle is dominated by the point \emph{F12.5\_A16} which has higher accuracy and lower current consumption.

\subsection{\textbf{Adaptive Low-Power Sensing Technique}}
\begin{figure}[t]
\vspace{-12pt}
\setlength{\abovecaptionskip}{0pt}
\setlength{\belowcaptionskip}{-12pt}
\centerline{\includegraphics[width=220pt,height=220pt,keepaspectratio]{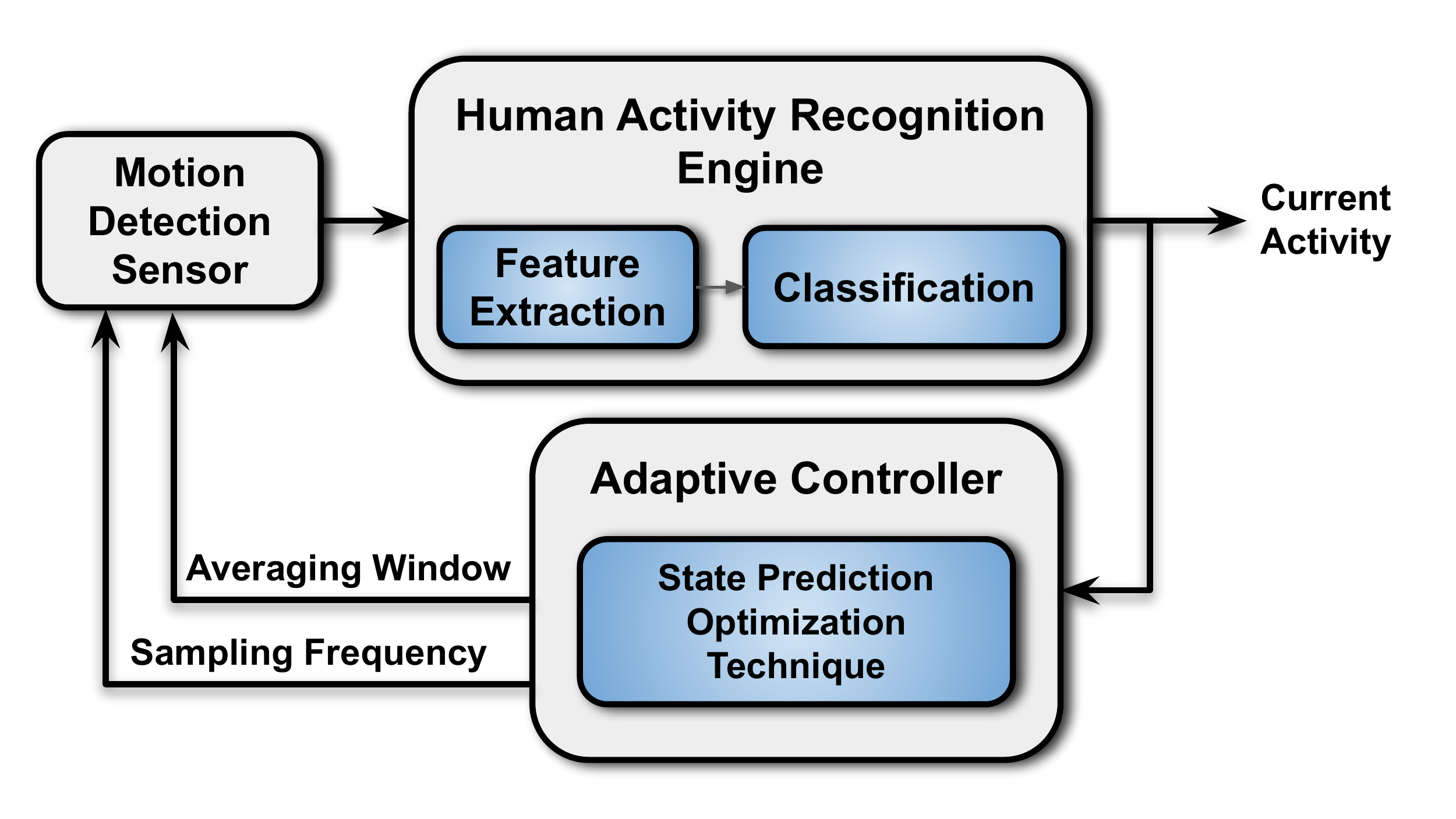}}
\caption{Low-Power Sensing HAR Framework.}
\label{lp_technique}
\vspace{-5pt}
\end{figure}

To reduce the sensor's power consumption, AdaSense introduces a new technique that switches among different sampling frequencies and averaging windows. Our adaptive controller dynamically switches to a lower power configuration when the user activity is stable (i.e, the user has been doing the same activity for a long time), and switches back to the highest accuracy configuration when the user activity changes to capture the right one. In other words, if the user has been walking for a certain period of time, it means that this user is steadily doing the same activity and will probably continue doing it for a while, so we can lower the sampling frequency and averaging window to reduce the power consumption of the sensor. However, if the activity keeps changing rapidly, then the sensor needs to operate at its highest power to capture the correct activity. Fig.~\ref{lp_technique} shows the proposed framework for AdaSense. The sensor first operates at its high power configurations, then it forwards the collected data to the HAR framework. As explained in Section ~\ref{sec:recognition_techniques}, the HAR framework classifies the data, and feeds its output to an adaptive controller which in turn adjusts the next episode's sensor configurations. Our controller uses a novel technique called state prediction optimization technique (SPOT) to take its decision.

\subsection{\textbf{The State Prediction Optimization Technique (SPOT)}}
The adaptive controller in AdaSense makes its decision regarding the sensor's configuration using our SPOT technique which is based on a simple finite state machine. First, we have four states representing the four optimal sensor configurations chosen during the design space exploration analysis earlier. These four states are sorted in descending order according to the power consumption. The accelerometer starts working at the first state \emph{(F100\_A128)}, and every one second it compares the current classification output with the previous classification output. When the activity stabilizes, i.e when the classifier output remains the same for few classification attempts called {\it stability threshold}, it moves to the lower power state. If the classified user activity changed at any state, the sensor returns back to the first high accuracy state.

\begin{figure}[t]
\setlength{\belowcaptionskip}{-2pt}
\setlength{\belowcaptionskip}{-15pt}
\centerline{\includegraphics[clip, trim=20pt 90pt 20pt 100pt, width=250pt,height=250pt,keepaspectratio]{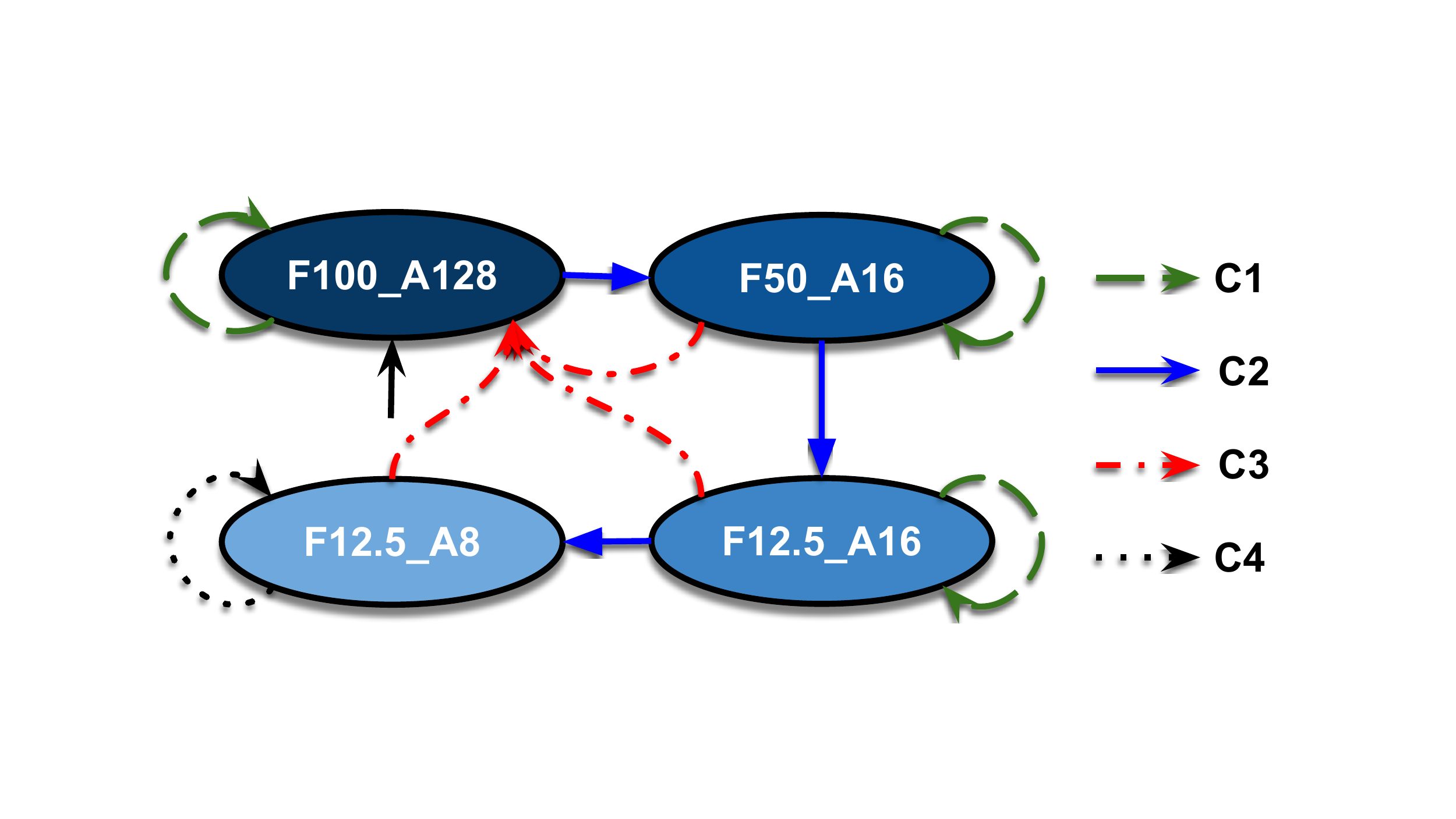}}
\caption{State Prediction Optimization technique FSM, states named as F(sampling frequency)\_A(averaging window), while C1-C4 are the conditions directing the flow from the first to the last state as a function of the activity stability.}
\label{fsm_spot}
\end{figure}

For example, suppose we chose to operate on four different states named \emph{\{F100\_A128, F50\_A16, F12.5\_A16, F12.5\_A8\}} as shown in Fig. \ref{fsm_spot}, and \emph{C1, C2, C3, C4} are the conditions that control the transition from one state to another such that:

\begin{itemize}
    \item \textbf{C1:} Current Activity == Last Activity \& \\*
    \hspace*{16pt} Counter ${<}$ stability threshold
    \item \textbf{C2:} Current Activity == Last Activity \& \\*
    \hspace*{16pt} Counter = stability threshold
    \item \textbf{C3:} Current Activity != Last Activity
    \item \textbf{C4:} Current Activity == Last Activity
\end{itemize}

\noindent SPOT starts by operating at \emph{F100\_A128}, and frequently classifies the output data from the sensor and increment a counter whenever the current activity matches the previous one. It continues to do so till the counter reaches the stability threshold. At that point, it switches to the next state in which it follows the same behaviour, successively switching to the next states till it reaches the last one and stays there. If at any state the current activity did not match the previous one, SPOT resets the counter, and switches immediately to the first high accuracy state, then the behaviour is repeated as long as the device is on.

\subsection{\textbf{The SPOT technique with confidence}}
The total power consumption depends on the time spent at each state. In SPOT, the decision to move from a lower power state to a higher power state is taken when the classifier reports a change in the human activity. The classifier reports that the activity is changed in two cases: when the activity actually changes, or when the classifier mispredicts due to some noise in the sensor's data. As a result, we introduce the {\it confidence} parameter in SPOT, which adds some tolerance to the noisy data. This confidence is the classifier's probability of the chosen output class; for example, if we have two output classes (walk, sit), and the softmax at the classifier's output layer gave the probabilities (0.8, 0.2), then the classification would be ``walk'' with confidence 0.8. In SPOT with confidence, the decision to move to a higher power state is taken when the classifier reports that the activity is changed with a confidence higher than a certain threshold called {\it confidence threshold}.

\vspace{-4pt}

\section{\textbf{Experimental Evaluation and Results}}
\label{sec:experiments}
\subsection{\textbf{Experimental Setup}}
\textbf{\emph{Hardware Used:}} We evaluate the proposed adaptive sensing technique using a Texas Instruments CC2640R2F MCU \cite{CC2640R2F} integrated with a Bosch Sensortec BMI160 16-bit inertial measurement unit (IMU) \cite{bmi160}. We operated the IMU in both the normal mode and the low-power mode to set different sampling frequencies and averaging windows.
 
\textbf{\emph{Data:}} We only enabled the IMU's accelerometer, and we collected data with the x, y and z sensor readings at different sampling frequencies and averaging windows. Then we used the collected data to evaluate our methodology. Using the setup mentioned in Section \ref{sec:recognition_techniques}, we trained our neural network on an extensive data set of 7300 activity windows of the four optimal acceleromenter configurations \emph{\{F100\_A128, F50\_A16, F12.5\_A16, F12.5\_A8\}} analyzed in table \ref{Freq_AVG_combinations}. The data recorded 6 different activities: walk, sit, lie down, go upstairs, going downstairs, stand.

\subsection{\textbf{AdaSense Behavioural Analysis}}
\begin{figure}[htbp]
 \centering
 \vspace*{-5pt}
  \begin{subfigure}{0.5\textwidth}
    \belowcaptionskip=0pt
    \abovecaptionskip=0pt
    \centering
    \includegraphics[width=240pt,height=240pt,keepaspectratio]{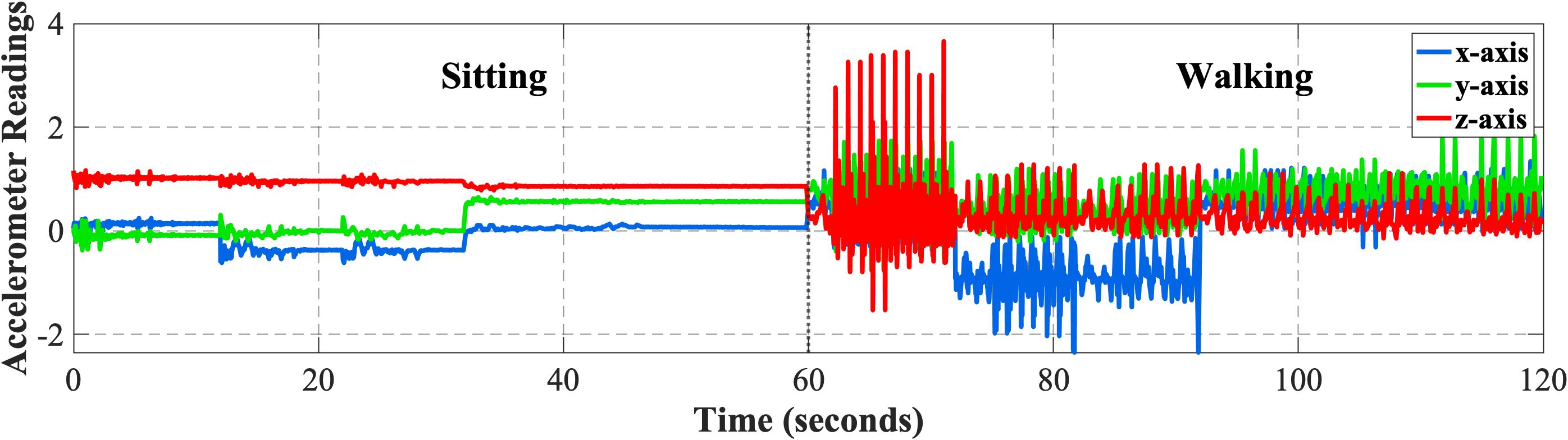}%
    \captionsetup{skip=0pt}
    \caption{3-axes Accelerometer Readings with Time.}
    \label{fig:demo1}
  \end{subfigure}

  \vspace*{4pt}

   \begin{subfigure}{0.5\textwidth}
    \belowcaptionskip=-5pt
    \abovecaptionskip=-5pt
    \centering
    \includegraphics[width=240pt,height=240pt,keepaspectratio]{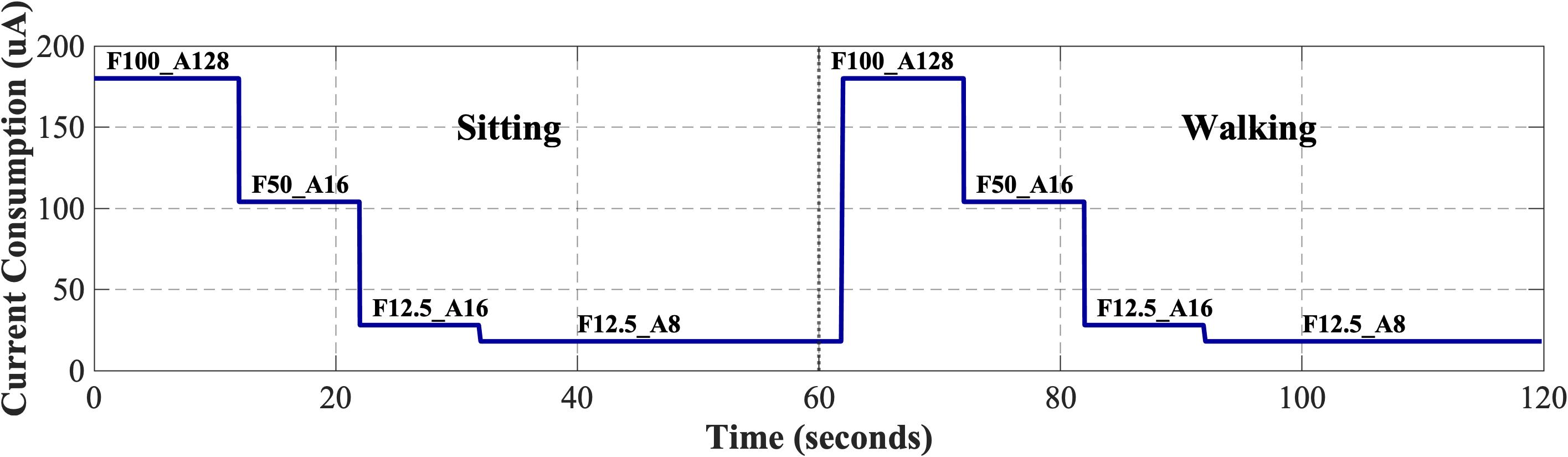}%
    \captionsetup{skip=0pt}
    \caption{Sensor Current Consumption per unit time.}
    \label{fig:demo2}
  \end{subfigure}
  \vspace*{4pt}
  \caption{AdaSense Behavioural Analysis.}
  \vspace*{-15pt}

  \label{fig:demoMain}
\end{figure}

As an illustration of AdaSense's performance, we analyze its inputs and outputs over a time interval of 120 seconds as shown in Fig. \ref{fig:demoMain}. We show a whole use case in which the user sits for the first 60 seconds, then the user changes the activity and starts to walk for another 60 seconds. Fig. \ref{fig:demo1} shows the inputs from the accelerometer over the chosen time interval where the y-axis is the x, y and z axes of the accelerometer reading. Fig. \ref{fig:demo2} shows the classification and power analysis, the y-axis is the current consumption in \si{\mu}A. We observe that AdaSense starts operating at the high power configuration of the sensor {\it F100\_A128}, and then it gradually switches to lower power configurations, until it reaches the minimum {\it F12.5\_A8} after 28 seconds. It stays there till the activity changes at time 60 seconds when the sensor switches back to the highest power state again, and repeatedly does the same behaviour till it also reaches the minimum after another 28 seconds.

\subsection{\textbf{Power \& Accuracy Analysis}}
\begin{figure}[t]
 \centering
 \vspace*{-5pt}
  \begin{subfigure}{0.5\textwidth}
    \belowcaptionskip=0pt
    \abovecaptionskip=0pt
    \centering
    \includegraphics[width=250pt,height=250pt,keepaspectratio]{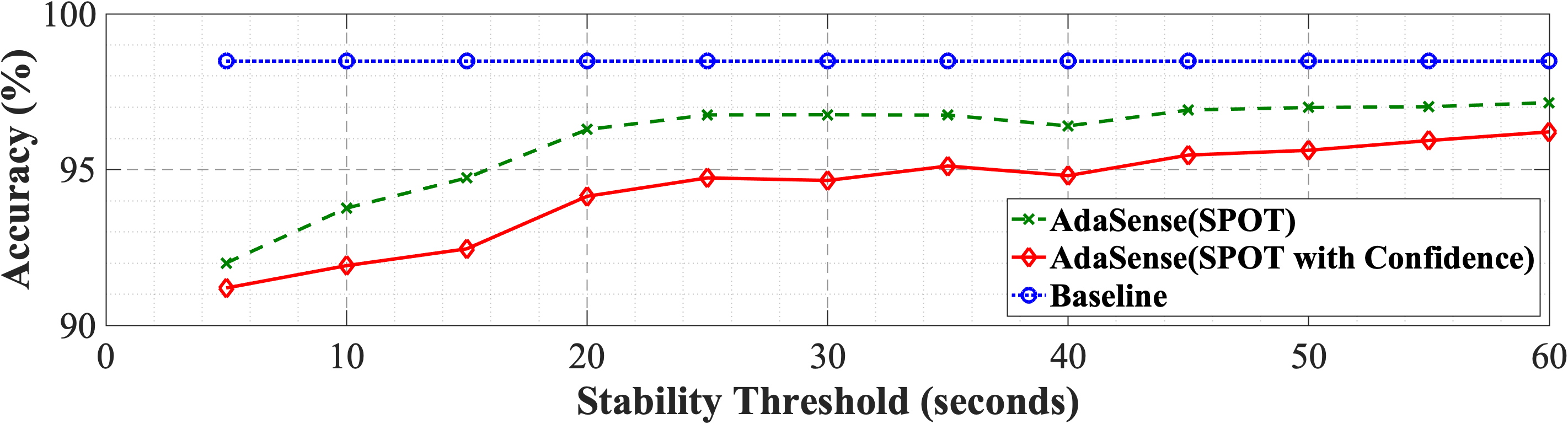}
    \captionsetup{skip=0pt}
    \caption{Classification Accuracy with Stability Threshold.}
    \label{accuracy_with_stability_threshold}
  \end{subfigure}

  \vspace*{4pt}

   \begin{subfigure}{0.5\textwidth}
     \belowcaptionskip=-5pt
    \centering
    \includegraphics[width=250pt,height=250pt,keepaspectratio]{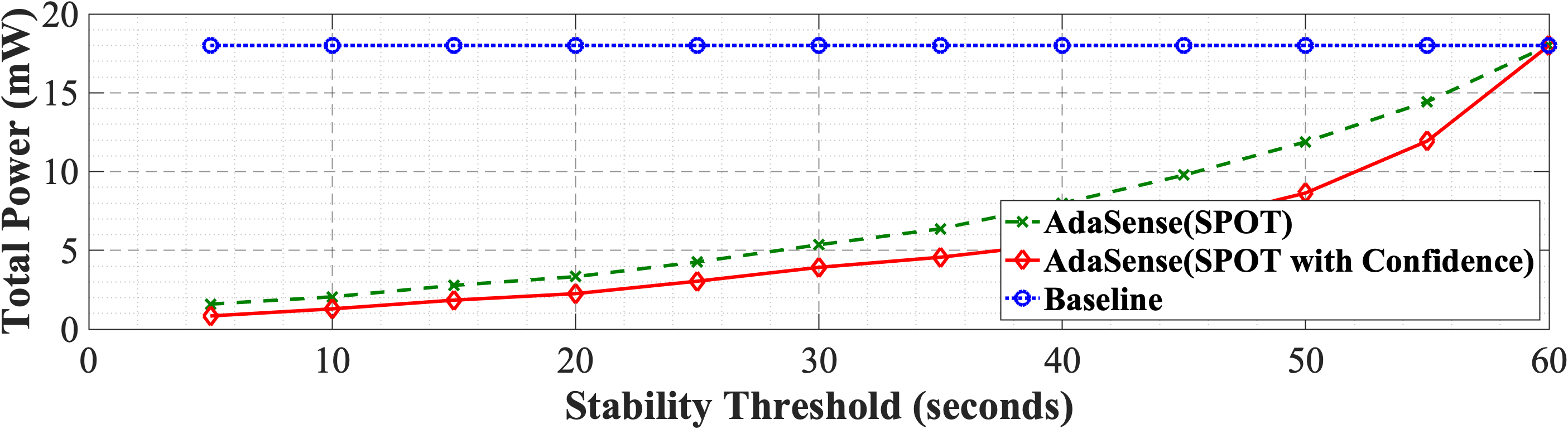}
    \caption{Total Power Consumption with Stability Threshold.}
    \label{power_with_stability_threshold}
  \end{subfigure}
  \vspace*{4pt}
  \caption{AdaSense Power and Accuracy analysis.}
  \label{adapt}
  \vspace*{-17pt}

\end{figure}

We first analyze the impact of the proposed methodology to co-optimize the sensor, feature extraction and classification of the HAR framework on the activity recognition accuracy and the sensor's power consumption. Fig.~\ref{accuracy_with_stability_threshold} shows the accuracy of classification as we increase the stability threshold which determines the stability of the activity; hence, switch the sensor to a lower configuration. We compare the accuracy under three scenarios. In the first scenario, we prevented the controller from switching among different sensor configurations, i.e. the sensor operates on the high power configuration \emph{\{F100\_A128\}} all the time; we take this as our baseline to measure how switching between different configurations affects the accuracy. In the second and the third scenarios, we analyze AdaSense using our two different adaptive controllers: SPOT and SPOT with confidence of value 0.85 respectively to switch among the four sensor configurations when the user activity becomes stable. In the three scenarios, we trained a single neural network on data from the four different accelerometer configurations mentioned before, and we used that model to test the classification accuracy while changing the setup of the adaptive sensor configurations controller.

From the results of Fig.~\ref{accuracy_with_stability_threshold}, we observe that the accuracy increases as the value of the stability threshold increases. Specifically, as the stability threshold increases from zero to 20 seconds, the classification accuracy rapidly increases from 91\% to 96.5\%. Then, the accuracy saturates within a range of 1.5\% below the baseline. This reduction in accuracy is expected because when the stability threshold is low (i.e., $<$ 20 seconds), the adaptive controller promptly switches to low-power accelerometer configurations; however, since the user activity is not stable for a long time, it triggers changes among different sensor configurations which result in a lower recognition accuracy. When the stability threshold is high enough (i.e., $>$ 20 seconds), the loss in accuracy is negligible when compared to the baseline.

Next, we compare the total power consumption for the sensor as a function of the stability threshold in seconds. As shown in Fig.~\ref{power_with_stability_threshold}, the power consumption increases with the increase in the stability threshold; when the stability threshold is low, the sensor rapidly switches to a low-power configuration after few seconds, so more time is spent at the lower power configurations which minimizes the total power consumption by the sensor. However, as the stability threshold increases, the time spent on the low-power sensor configuration decreases, so the total power increases. Furthermore, at stability threshold of 60 seconds, the power consumption matches the baseline as the sensor spends all the time operating at the high power configuration. In average, the total power can be reduced by 60\% using SPOT and 69\% using SPOT with confidence.

\subsection{\textbf{Comparison to the previous work}}
\begin{figure}[t]
\centerline{\includegraphics[clip, trim=0pt 0pt 0pt 50pt, width=250pt,height=250pt,keepaspectratio]{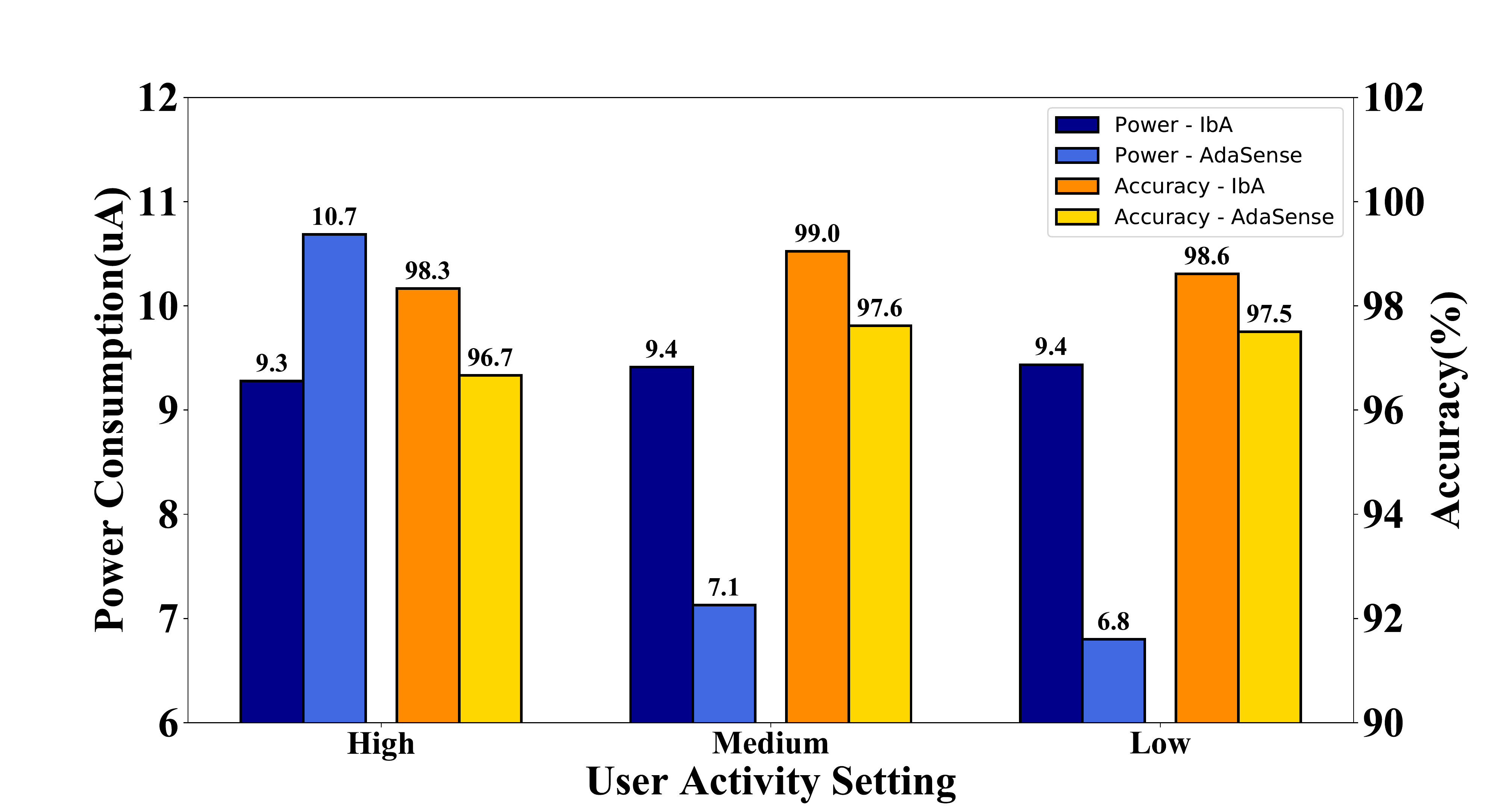}}
\caption{Comparison between AdaSense and Intensity Based Approach(IbA) \cite{8782506} in terms of Accuracy and Power Consumption under different user activity settings.}
\label{spot_vs_existing_power}
\vspace*{-20pt}

\end{figure}

This section compares the accuracy and the power consumption of AdaSense to the related work. NK {\it et al.} \cite{8782506} use an activity intensity based approach to reduce the power consumption of the sensors; the sensors switch to low-power mode with low-intensity user activities (i.e. stand, sit, lie down), and operate at the normal mode with more intense activities (i.e. walk, go upstairs, go downstairs). NK {\it et al.} define the intensity of the activity using the first derivative of the accelerometer readings, and they retrain separate classifiers for each used sampling frequency. Fig. \ref{spot_vs_existing_power} summarizes the comparison, the x-axis shows three user activity settings \{{\it High, Medium, Low}\}, which differs in terms of the user activity change rate. {\it High} means that the user activity is not stable (i.e. changes every 10 seconds), while {\it Low} means that the user activity is quite stable (i.e. it takes the user at least 1 minute to change the activity). The left y-axis shows the power consumption of using AdaSense versus the activity intensity based technique. As expected, when the user activity setting is high, AdaSense spends most of the time at the high power sensor configuration; therefore, the power consumption is relatively high. However when the user activity starts to be more typical, the power consumption is reduced by at least 25\% compared to the previous work. The right y-axis shows the recognition accuracy. The results show that AdaSense has has slightly lower recognition accuracy (i.e ranging from 1\% to 1.5\%) depending on the setting in comparison to the technique used by NK {\it et al.}. This loss in accuracy is acceptable in trade of the significant power and memory savings.

\noindent \textbf{Memory Requirements:} While NK {\it et al.} retrain different neural networks for the different sampling frequencies, AdaSense trains a single classifier on data from different sensor configurations, consuming 2$\times$ less memory to store the classifier(s) weights. This memory reduction is important for wearable devices as they only have few KBs of memory.

\noindent \textbf{Data Processing Overhead:} In AdaSense, we do not need to compute the derivative of the collected sensor data to switch among the different configurations. Therefore, we prevent computations overhead that might compromise the power savings from the sensor.

\section{\textbf{Conclusion}}
\label{sec:conclusion}

Wearable devices have many advantageous applications in health services, and with the advancement in the research done to reduce the power consumption on those wearable devices, their deployment in real applications would become more practical. This paper presented a low-power sensing technique for activity recognition on wearable devices. Using an adaptive controller, we dynamically switched the sensor to lower sampling frequencies and averaging windows depending on the stability of the user-activity. We analyzed the trade-off between the accuracy and the power consumption for different sensor configurations. Then, we designed an adaptive controller that switches among the resulting optimal sensor configurations. We also co-optimized the features extraction and the classification achieving up to 69\% reduction in total power consumption with less than 1.5\% degradation in the recognition accuracy.

\vspace{2pt}
\noindent \textbf{Acknowledgements:} This work is partially supported by NSF grant 1814920 and DoD ARO grant W911NF-19-1-0484.

\bibliographystyle{IEEEtran}
\bibliography{refbib}
\end{document}